\documentclass[trackchanges,twocolumn]{aastex701}

\begin{document}

\title{Relativistic Effects on Circumbinary Orbit Stability}

\author[0000-0002-2780-264X]{Gonzalo C. de Elía}
\affiliation{Instituto de Astrofísica de La Plata, CCT La Plata-CONICET-UNLP, Paseo del Bosque S/N (1900), La Plata, Buenos Aires, Argentina}
\affiliation{Facultad de Ciencias Astronómicas y Geofísicas de La Plata, UNLP, Paseo del Bosque S/N (1900), La Plata, Buenos Aires, Argentina}
\email[show]{gdeelia@fcaglp.unlp.edu.ar}

\author[0009-0009-0042-638X]{Macarena Zanardi}
\affiliation{Instituto de Astrofísica de La Plata, CCT La Plata-CONICET-UNLP, Paseo del Bosque S/N (1900), La Plata, Buenos Aires, Argentina}
\affiliation{Facultad de Ciencias Astronómicas y Geofísicas de La Plata, UNLP,
Paseo del Bosque S/N (1900), La Plata, Buenos Aires, Argentina}
\email{mzanardi@fcaglp.unlp.edu.ar}

\author[0000-0003-2401-7168]{Rebecca G. Martin}
\affiliation{Nevada Center for Astrophysics, University of Nevada, Las Vegas, 4505 South Maryland Parkway, Las Vegas, NV 89154, USA}
\affiliation{Department of Physics and Astronomy, University of Nevada, Las Vegas, 4505 South Maryland Parkway, Las Vegas, NV 89154, USA}
\email{rebecca.martin@unlv.edu}

\begin{abstract}

With $n$-body simulations and analytic approximations we study the dynamics and stability of low eccentricity misaligned test particles around binary systems with varying mass fraction and eccentricity. General relativity (GR) plays a primary role in determining the motion  of an outer particle since it drives apsidal precession of the binary orbit. The effects of GR can drive particle instability close to the binary orbit, depending upon the binary parameters and the initial inclination of the particle. For the binary parameters we consider, we find instability up to a semimajor axis of about $8\,a_{\rm b}$, where $a_{\rm b}$ is the binary semimajor axis. In particular, we identify and analyse three different regions of instability that are driven by GR in the phase plane of the initial semimajor axis and the initial inclination of the particle. The results have implications for circumbinary orbits and circumbinary disks on all scales, but are particularly important around supermassive black hole binaries where the effects of GR can be strong. 

\end{abstract}


\keywords{\uat{Binary stars}{154} --- \uat{Black holes}{162} --- \uat{Dynamical evolution}{421} --- \uat{General relativity}{641} --- \uat{N-body simulations}{1083} --- 
\uat{Supermassive black holes}{1663} }

\section{Introduction}
\label{sec:intro}

Circumbinary objects and disks are observed on a wide range of scales in the universe from the moons around the Pluto-Charon binary in the solar system \citep{Youdin2012}, to circumbinary planets \citep[e.g.][]{Doyle2011,Welsh2012} and disks \citep[e.g.][]{Chiang2004}, to stars and disks around supermassive black hole (SMBH) binaries in the centers of AGN \citep{DeRosa2019,Agazie2023}. Misaligned circumbinary disks may form from chaotic accretion processes around stellar binaries \citep{Chiang2004,Capelo2012,Brinch2016} and SMBH binaries \citep{King2006,Li2022}. Circumbinary objects may be found in an orbit that is misaligned to the binary orbit if they formed in a misaligned disk or if they were captured \citep[e.g.][]{Valtonen2006}.

Understanding the stability of misaligned test particle orbits around binaries is important both in terms of the stability of a low mass object, but also understanding the evolution of a circumbinary accretion disk. Each ring of a low mass disk feels the same torque as the particle. The disk rings communicate with each other either through viscous communication or wave-like communication and the disk can undergo global nodal precession \citep[e.g.][]{Larwood1996,Martin2017}. If the communication is too slow, this can lead to disk breaking where multiple disk rings are formed that can undergo nodal precession on different timescales \citep[e.g.][]{Nixon2012,Aly2015}. The inner truncation radius of a circumbinary accretion disk is also close to the inner radius that stable test particle orbits exist \citep[e.g.][]{Artymowicz1994}.

There are two types of nodal precession that a test particle orbit may undergo \citep[e.g.][]{Ziglin1975,Farago2010,Naoz2017,Zanardi2017,Chen2019}. First, it may undergo {\it circulating} nodal precession where the angular momentum vector of the particle precesses about the binary angular momentum vector. Second, it may undergo {\it librating} nodal precession where the particle precesses about a stationary inclination. In the absence of general relativity (GR), the stationary inclination is aligned to the binary eccentricity vector. In this case,  the nodal precession type of a test particle is determined by the binary eccentricity and the initial particle inclination. There is no dependence on the orbital semimajor axis of the particle.

While the stability of test particle orbits around a binary have been explored in detail \citep{Verrier2009,Chen2020}, the effects of GR have most often been assumed to be small. GR causes prograde apsidal precession of a binary orbit \citep{Einstein1915}. This leads to a radial dependence of the stationary inclination \citep{Zanardi2018}.  Close to the binary, circumbinary test particle orbits may still undergo nodal libration if their precession timescale is shorter than the binary apsidal precession timescale \citep{Childs2021}. There is a critical semimajor axis, $a_{\rm lim}$, outside of which only circulating nodal precession occurs \citep{Lepp2022, Zanardi2023}. Inside of this semimajor axis, the particle may undergo nodal circulation or nodal libration depending upon its initial inclination.

The value of $a_{\text{lim}}$ can be assumed to be a measure of the importance of GR effects. Small values of $a_{\text{lim}}$ lead to significant changes in the orbital inclinations that define the motion regimes of particles close to the binary compared to those obtained without GR. The limits of the motion regimes of a particle only show drastic changes farther away from the binary for large values of $a_{\text{lim}}$. The parameter $a_{\text{lim}}$ is an increasing function of the binary separation. For stellar mass binaries, small values of $a_{\text{lim}}$ require sufficiently small binary separations so that tidal effects likely lead to circularization of the binary. However, SMBH binaries may remain eccentric even at separations where $a_{\text{lim}}$ reaches small values
\citep{Childs2024}, at least until gravitational waves (GWs) take over \citep{Begelman1980}. Observations of merging SMBHs suggest that some are still eccentric \citep[e.g.][]{Jiang2022,Gualandris2022}.

In this paper, we explore the dynamics of test particles around a binary orbit including the effects of GR. In Section~\ref{sec:section1} we describe the simulation set-up. In Section~\ref{sec:section2} we present survival maps with and without GR. We find that GR can lead to particle orbit instability close to the binary orbit. In Section~\ref{sec:section3} examine in more detail three regions of parameter space that become unstable through GR. We discuss the variation of the results with the simulation parameters in Section~\ref{sec:seccion_testeo_parametros}. Finally, Section~\ref{sec:section4} presents the conclusions of our research.


\section{Simulation set-up} \label{sec:section1}

We study the dynamical properties and stability of outer massless particles orbiting around a binary system including GR. To do this, we make use of IAS15, a 15th-order integrator with adaptive step-size control in the {\sc rebound} $n$-body code \citep{Rein2015}\footnote{While numerous authors have adopted the WHFast integrator to study the dynamics of circumbinary orbits \citep[e.g.][]{Chen2020, Abod2022, Lepp2022, Martin2022, Childs2023}, we decide to use IAS15. On the one hand, WHFast is very fast and provides an accurate implementation of a Wisdom-Holman symplectic integrator for long-term orbit integrations of planetary systems \citep{Rein2015_whfast}, being the best choice for systems with a dominant central object and small perturbations to Keplerian orbits. On the other hand, IAS15 is a fast, very high-order, non-symplectic integrator, which is ideal for accurately describing the dynamics around binary systems.}. In particular, GR effects are modeled using the $\text{grfull}$ package included in {\sc rebound}x, which incorporates first-order post-Newtonian effects from all bodies in the system \citep{Tamayo2020}. 

The binary system has a total mass $m_{\text{b}}$ = $m_1$ + $m_2$, where $m_1$ and $m_2$ are the masses of the individual binary components, with $m_2 \leq m_1$. The binary orbit has semimajor axis $a_{\text{b}}$, eccentricity $e_{\text{b}}$, inclination $i_{\text{b}}$, longitude of ascending node $\Omega_{\text{b}}$, argument of pericenter $\omega_{\text{b}}$, and  true anomaly $\nu_{\text{b}}$. The mass fraction of the binary system is $f_{\text{b}} = m_{2}/m_{\text{b}}$.

The orbit of the outer massless test particle is described relative to a coordinate system whose origin is at the center of mass of the binary, the $xy$ plane coincides with the invariant plane of the system, and the $x$-axis is oriented toward a fixed reference direction. The six elements that define the orbit of the outer test particle are the semimajor axis $a$, the eccentricity $e$, the inclination $i$, the longitude of ascending node $\Omega$, the argument of pericenter $\omega$, and the true anomaly $\nu$. 
The definition of the nodal libration and nodal circulation regimes of a test particle due to GR requires that $\Omega$ be measured relative to the pericenter of the binary, so we will be interested in analyzing the behavior of $\Omega$ - $\omega_{\text{b}}$ \citep{Zanardi2018}.

\begin{figure*}[ht!]
\centering
\includegraphics[angle=270, width=0.98\textwidth]{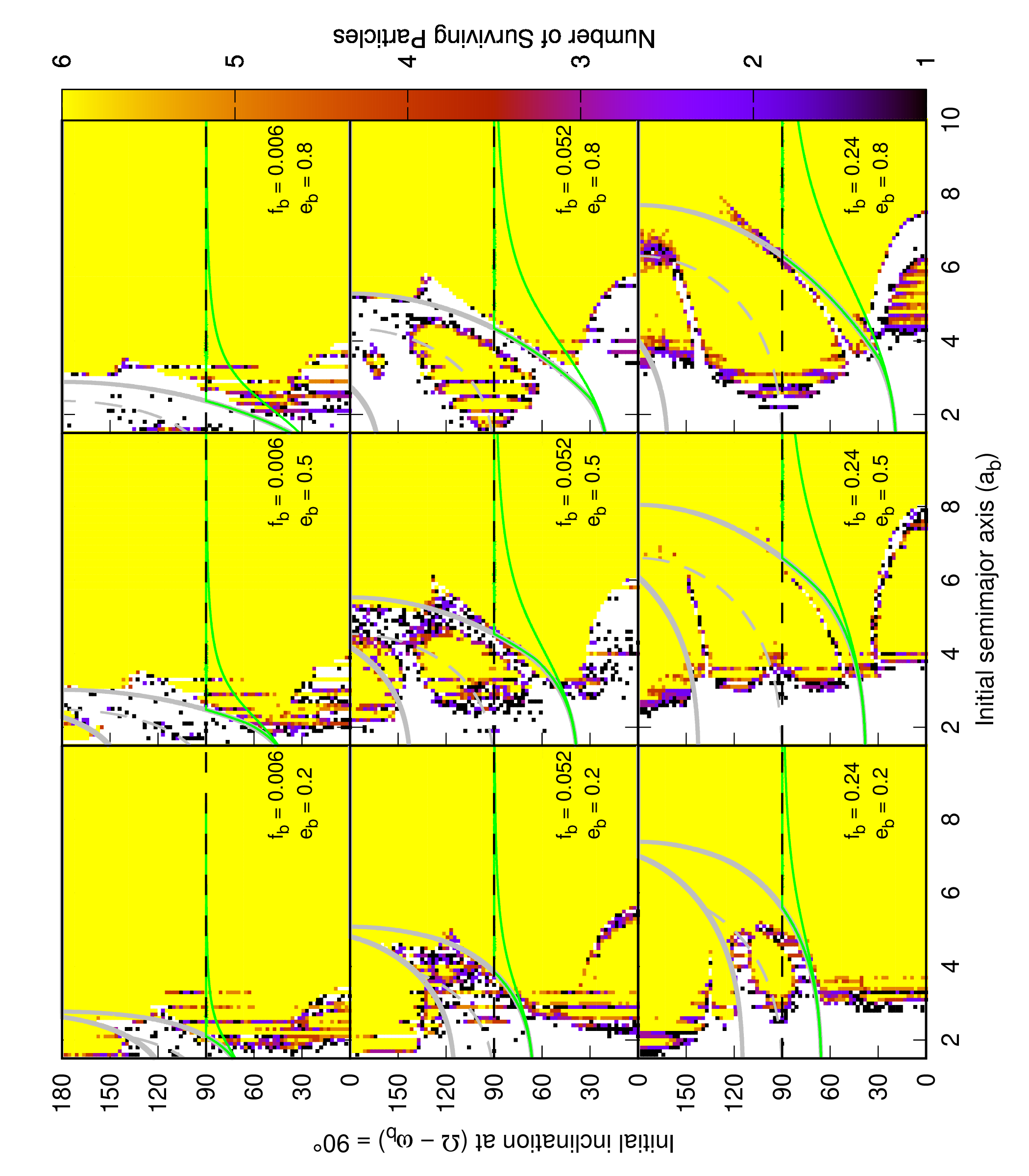}
\caption{Survival maps in the plane  ($a_{\text{ini}}$, $i^{\text{e}}_{\text{ini}}$) considering GR effects. From left to right, the columns are associated with $e_{\text{b}}$ = 0.2, 0.5, and 0.8. From top to bottom, the rows correspond to $f_{\text{b}}$ = 0.006, 0.052, and 0.24. In each panel, the solid gray curves define the extreme inclinations of the nodal libration region from a secular theory up to the quadrupole level. The dashed gray curves within the nodal libration region separate the initial conditions that produce orbital flips from those that lead to purely retrograde orbits. The green curves delimit the nodal circulation region with orbital flips. The color code illustrates the number of surviving particles after 1 Myr. The white areas are regions where no particles survive.
\label{fig:mapa_gr}}
\end{figure*}


Our research is aimed at analyzing the role of the GR in the dynamical properties and stability of outer test particles on orbits with a low initial eccentricity around a binary. 
GR constrains the semimajor axis of an outer test particle that experiences nodal libration around an inner binary to $a <a_{\rm lim}$ \citep{Lepp2022,Zanardi2023}. The upper limit of the semimajor axis, $a_{\text{lim}}$, is given by
\begin{eqnarray}
a_{\text{lim}} = \left( \frac{m_{1}f_{\text{b}}c^{2}(1-e^{2}_{\text{b}}) (1 + 4 e^{2}_{\text{b}})} {4 k^{2} m^{2}_{\text{b}} (1 - e^{2})^{2}} a^{9/2}_{\text{b}}  \right)^{2/7},
 \label{eq:alim}
\end{eqnarray}
where $c$ is the speed of light and $k^{2}$ is the gravitational constant.


The dynamics of circumbinary orbits depends strongly upon $a_{\rm lim}$. If $a_{\rm lim} \sim a_{\rm b}$, then the orbits of the particles can only be in the nodal circulation regime. The apsidal precession of the binary driven by GR is so strong that the outer particles see an effectively circular orbit binary. If $a_{\rm lim} \gg a_{\rm b}$, then particle orbits close to the binary may be found in the nodal libration regime, depending upon their initial inclination. In this case, the effects of GR are weak and particle stability may be similar to the case without GR \citep[see e.g.][]{Doolin2011, Chen2020}.

We choose the binary parameters such that GR is important. From this, we select a total binary mass $m_{\text{b}} = 10^{7}\,\rm  M_{\odot}$ and a semimajor axis $a_{\text{b}} =$ 0.01 pc for all our scenarios of study \citep[c.f.][]{Childs2024}. These choices are motivated through observations. On the one hand, SMBHs have masses in the range $10^6-10^{10}\,\rm M_\odot$. On the other hand, the semimajor axis can be down to about $0.01\,\rm pc$ before GWs become important \citep{Peters1964,Nixon2011}. For example, the SMBH binary candidate in the radio-quiet quasar SDSS J0159+0105 has an estimated total mass of 1.3 $\times$ 10$^{8}$ M$_{\odot}$ and a separation of 0.013 pc \citep{Zheng2016}. We choose a lower mass than this observed system to weaken the effects of GR so that our values for $a_{\rm lim}$ are in the interesting range (see Eq.~\ref{eq:alim}).

The SMBH binary adopts an inclination $i_{\text{b}} =0^{\circ}$, a longitude of ascending node $\Omega_{\text{b}} = 0^{\circ}$, an initial argument of pericenter $\omega_{\text{b}} = 0^{\circ}$, and an initial true anomaly $\nu_{\text{b}}$ = 180$^{\circ}$.
 We consider $f_{\text{b}}$ and $e_{\text{b}}$ as variables of interest of our study. The default parameters are $f_{\text{b}} =$ 0.052 and $e_{\text{b}} =$ 0.5, which have a critical semimajor axis $a_{\text{lim}}= 5.5 \,a_{\text{b}}$. From Eq.~\ref{eq:alim}, the critical semimajor axis $a_{\text{lim}}$ is an increasing function of the binary mass fraction $f_{\text{b}}$. Thus, we also explore the dynamics in systems with a value of $f_{\text{b}}= 0.006$ and $f_{\text{b}} = 0.24$, which lead to values of $a_{\text{lim}}$ of $3\, a_{\text{b}}$ and $8 \,a_{\text{b}}$, respectively, for $e_{\text{b}} = 0.5$.  

\begin{figure*}[ht!]
\centering
\includegraphics[angle=270, width=0.98\textwidth]{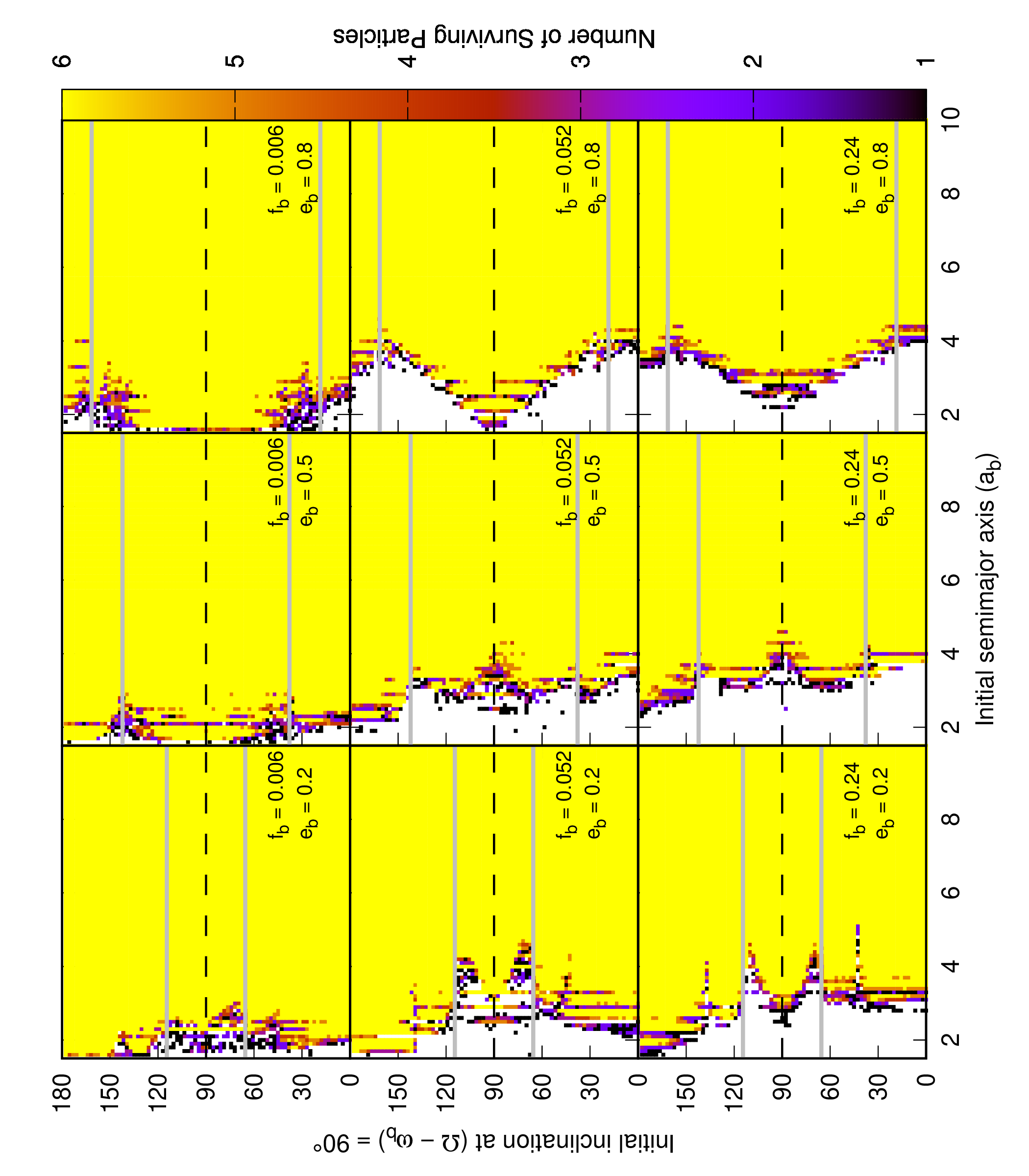}
\caption{As in Fig.\ref{fig:mapa_gr}, except that GR effects are not included. Note that in this case, the solid gray lines that define the extreme inclinations of the nodal libration region lead to constant values in each panel since they only depend on $e_{\text{b}}$.
\label{fig:mapa_singr}}
\end{figure*}

\begin{figure}[ht!]
\includegraphics[angle=270, width=0.5\textwidth]{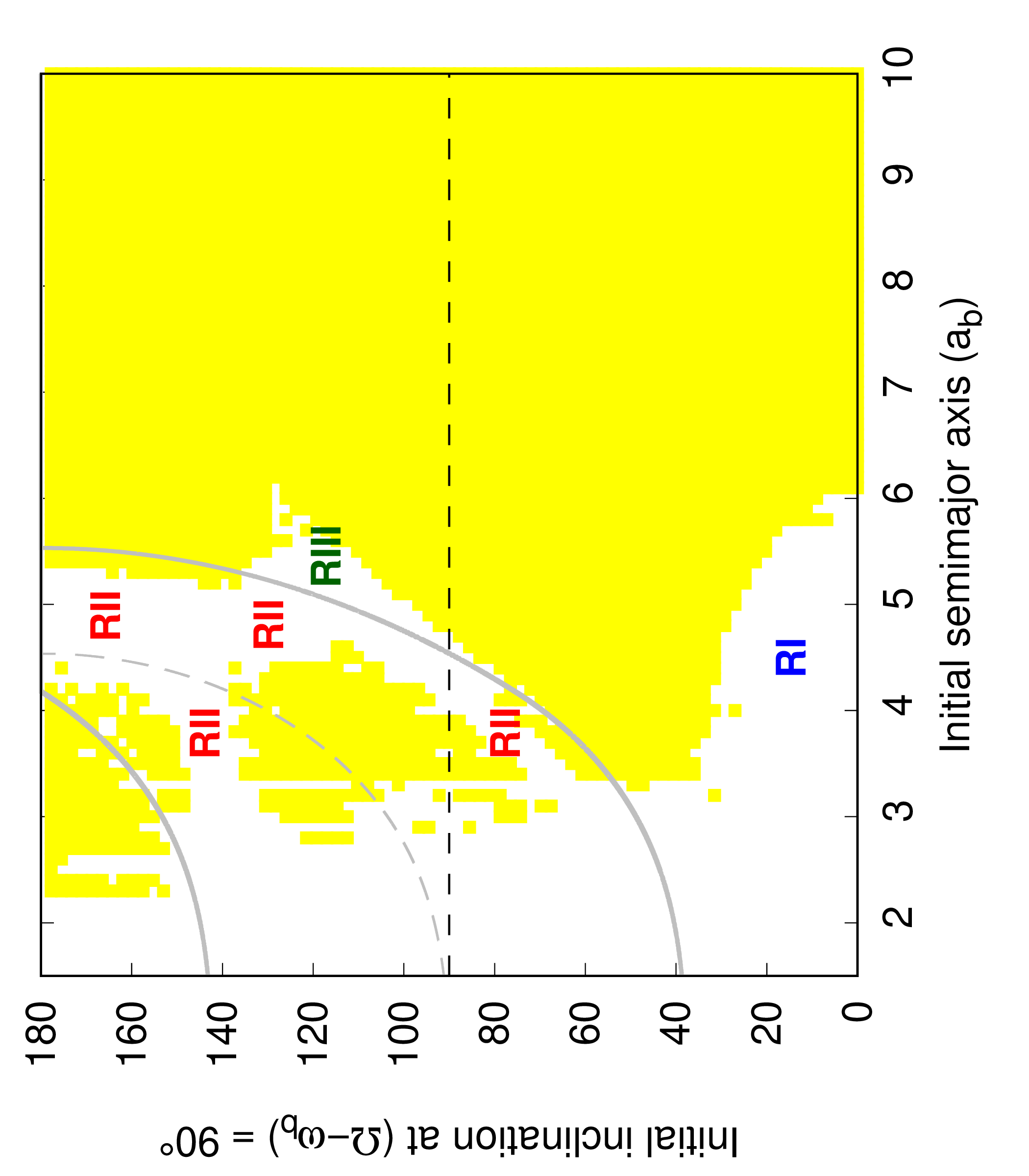}
\caption{Strongly unstable RI, RII, and RIII regions driven by GR in our default scenario. The yellow and white areas represent the stable and unstable regions, respectively. The references associated with the solid and dashed gray curves are given in the caption of Fig.~\ref{fig:mapa_gr}.
\label{fig:ejemplo_regiones}}
\end{figure}

Beyond the default case that assumes $e_{\text{b}} = 0.5$, we also consider systems with $e_{\text{b}} = 0.2$ and $e_{\text{b}} = 0.8$ for each of the three values of $f_{\text{b}}$ previously specified. The critical value $a_{\text{lim}}$ does not depend strongly on the binary eccentricity, except for extremely large values of $e_{\text{b}}$ \citep{Lepp2022,Zanardi2023} and $a_{\text{lim}}$  remains close to $3 \,a_{\text{b}}$, $5.5 \,a_{\text{b}}$, and $8 \,a_{\text{b}}$ for $f_{\text{b}} = 0.006$, 0.052, and 0.24, respectively, for the three chosen values of $e_{\text{b}}$.

\begin{figure*}[ht!]
\includegraphics[width=0.98\textwidth]{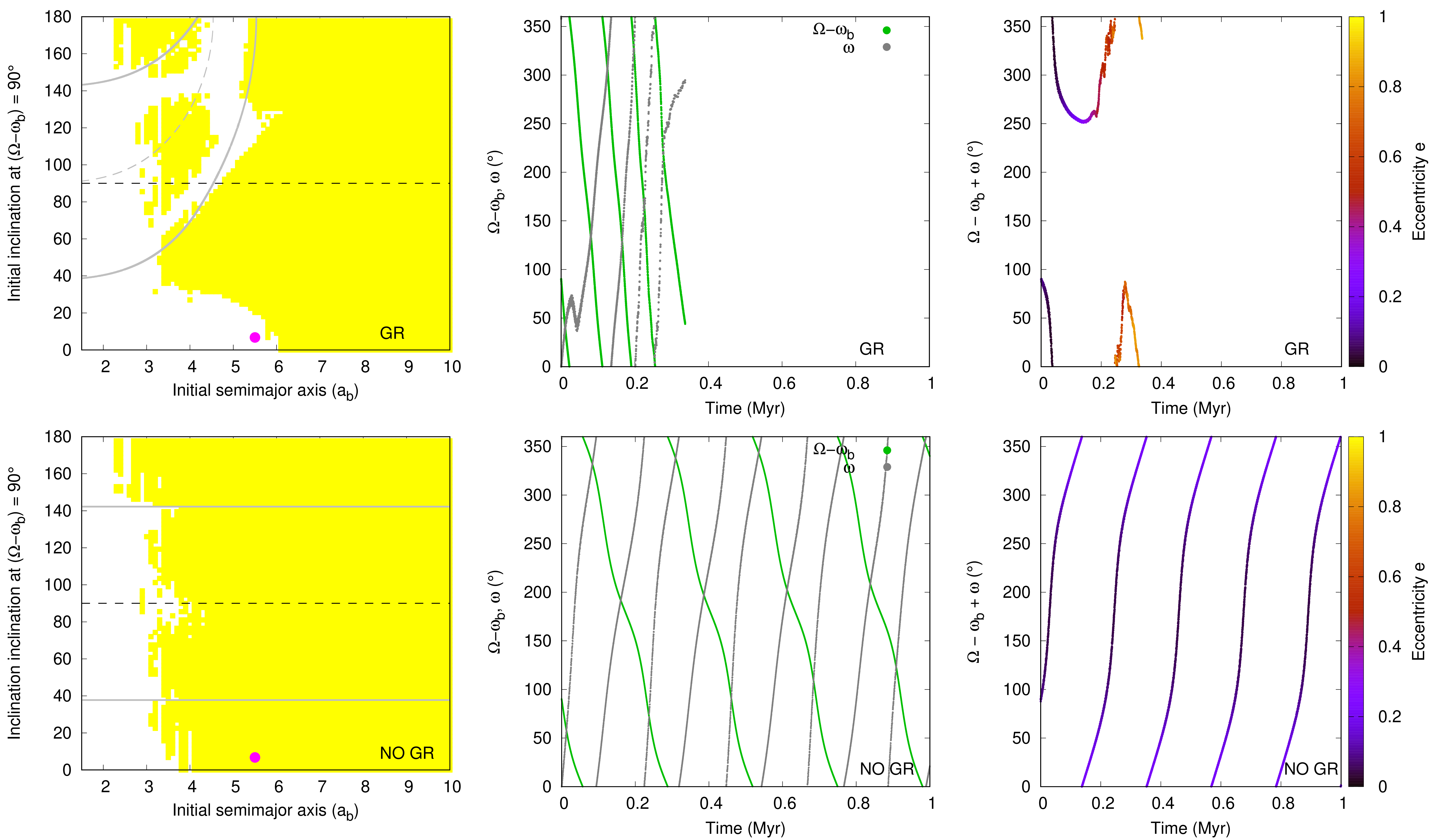}
\caption{Dynamical analysis of a test particle belonging to the RI region with (top panels) and without (bottom panels) GR effects for our default scenario. Left panels: The pink point represents the initial values of $a$ and $i^{\text{e}}$ of the particle. The stable and unstable regions are illustrated in yellow and white, respectively. The solid and dashed gray curves in the top panel and the solid gray lines in the bottom panel are referenced in the captions of Figs.~\ref{fig:mapa_gr} and \ref{fig:mapa_singr}, respectively. Middle panels: Temporal evolution of $\Omega - \omega_{\text{b}}$ (green points) and $\omega$ (gray points). Right panels: Temporal evolution of $\Omega - \omega_{\text{b}} + \omega$, where the color code illustrates the eccentricity of the test particle.  
\label{fig:ejemplo_region1}}
\end{figure*}

The three values specified for both $f_{\text{b}}$ and $e_{\text{b}}$ define the nine systems of work of our investigation. To study the stability of an outer test particle around a binary in each of those systems, we explore a wide range of values associated with its initial orbital parameters. First, the initial value of the semimajor axis $a$ is defined between 1.5$\,a_{\text{b}}$ and 10$\,a_{\text{b}}$ with a spacing of 0.1$\,a_{\text{b}}$. Then, the initial value of the inclination $i$ ranges from 0$^{\circ}$ to 180$^{\circ}$ with a spacing of 2.25$^{\circ}$\footnote{The existence of retrograde circumbinary orbits can be justified. According to \cite{Zanardi2018}, the combined action between the gravitational influence of the binary and the GR effects can produce retrograde inclinations from prograde initial values when the outer particle undergoes orbital flips for a limited range of parameter space in both the nodal libration and nodal circulation regimes. Moreover, the generation of retrograde structures around SMBHs could be the natural result of a series of small-scale, randomly oriented accretion events \citep[e.g.][]{King2006}. Retrograde circumbinary disks have also been recently observed around young stellar binaries \citep{Chen2026}.}, while the initial longitude of ascending node $\Omega$ and the argument of pericenter $\omega$ are always assumed to be equal to 90$^{\circ}$ and 0$^{\circ}$, respectively. It is worth noting that an inclination $i$ at $\Omega - \omega_{\text{b}} = 90^{\circ}$, which is called $i^{\text{e}}$, is associated with an extreme value in the evolutionary trajectory of an outer test particle, according to the secular quadrupolar theory without and with GR effects developed by \citet{Ziglin1975} and \citet{Zanardi2018}, respectively. Finally, for each set $a$, $e$, $i$, $\Omega$, and $\omega$, the true anomaly $\nu$ ranges from 0$^{\circ}$ to 300$^{\circ}$ adopting a spacing of 60$^{\circ}$. 
The outer test particles have an initial eccentricity $e =0.1$.

We analyze the orbital evolution of the outer test particles including GR effects over a time of $1 \,\rm Myr$, which is about $3.4 \times 10^{4}$ binary orbital periods. For our study on stability, a given test particle is removed from the system if $e >1$, $a < a_{\text{b}}$, or $a > a_{\text{oc}}$ = 20 $\, a_{\text{b}}$ during its orbital evolution, where $a_{\text{oc}}$ is the semimajor axis associated with the outer cutoff \citep[e.g.][]{Quarles2020,Chen2020}. Finally, for comparison, we also study the dynamics of outer test particles over $1\,\rm  Myr$ in the nine binary systems without the effects of GR. Such a comparative analysis allows us to improve our understanding concerning the role of GR in our systems of interest. 
We remark that the values adopted for the physical and orbital parameters of the system are those specified in this section unless explicitly stated otherwise.


\begin{figure*}[ht!]
\includegraphics[width=0.98\textwidth]{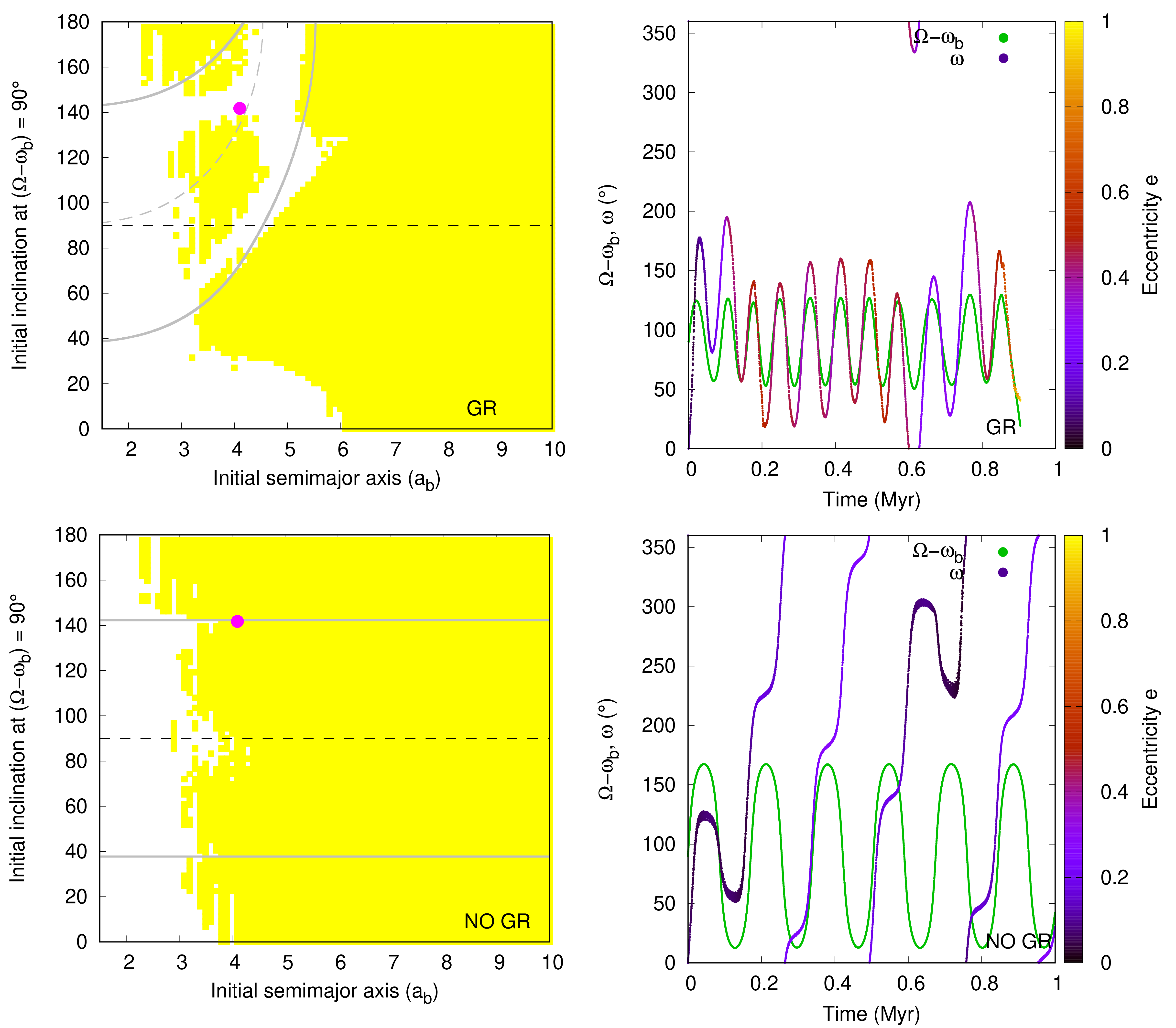}
\caption{Dynamical analysis of a test particle belonging to the RII region with an initial retrograde $i^{\text{e}}$ with (top panels) and without (bottom panels) GR effects for our default scenario. Left panels: As in the left panels of Fig.~\ref{fig:ejemplo_region1}. Right panels: Temporal evolution of $\Omega - \omega_{\text{b}}$ (green points) and $\omega$ (color points), where the color code represent the eccentricity of the test particle.}
\label{fig:ejemplo_region2_1}
\end{figure*}

\section{Survival maps} 
\label{sec:section2}

Fig.~\ref{fig:mapa_gr} shows survival maps of the test particles in the plane ($a_{\text{ini}}$, $i^{\text{e}}_{\text{ini}}$) for the nine work scenarios of our study that include GR effects, where $a_{\text{ini}}$ and $i^{\text{e}}_{\text{ini}}$ represent initial values of $a$ and $i^{\text{e}}$, respectively.
The left, middle, and right columns are associated with a value of $e_{\text{b}}$ = 0.2, 0.5, and 0.8, respectively. The top, middle, and bottom rows refer to a value of $f_{\text{b}}$ = 0.006, 0.052, and 0.24, respectively. 

Each pixel corresponds to six simulations, each with a different initial true anomaly. The color code corresponds to the number of particles that survive at the end of the simulation.  The white areas represent strongly unstable regions where no particles survive. The yellow areas represent stable regions, since all particles survive until the end of the simulation. The remaining areas of the map are assumed to be unstable regions, since at least one particle is removed from the system, and those that survive can show significant changes in their final orbital elements relative to their initial values.

The secular quadrupole theory including GR  \citep{Zanardi2018,Zanardi2023} allows us to specify the conditions that define the different regimes of motion of an outer test particle. The two gray solid curves show the extreme inclinations that define where a test particle undergoes nodal libration. A particle with an initial inclination between these two curves undergoes nodal libration, otherwise, it undergoes nodal circulation. The test particle can experience two different regimes of motion within the nodal libration region. On the one hand, prograde inclinations and retrograde inclinations above the gray dashed curve and below the upper gray solid curve cause nodal librations with orbital flips, meaning that the orbit is sometimes closer to prograde and sometimes closer to retrograde relative to the binary orbit. On the other hand, retrograde inclinations below the gray dashed curve lead to nodal librations with purely retrograde orbits. Outside the region delimited by the gray solid curves, the particles undergo nodal circulation. In addition, particles that begin between the green curves undergo nodal circulation with orbital flips. Otherwise, the particles undergo nodal circulation with purely prograde or purely retrograde orbits for values of $i^{\text{e}}_{\text{ini}}$ less than or greater than 90$^{\circ}$, respectively.

Fig.~\ref{fig:mapa_singr} shows survival maps of the test particles in the plane ($a_{\text{ini}}$, $i^{\text{e}}_{\text{ini}}$) for our nine working scenarios in the absence of GR. These are in agreement with previous simulations of coplanar \citep[e.g.][]{Holman1999,Sutherland2016} and misaligned particles \citep[e.g.][]{Doolin2011,Quarles2016,Hong2019,Chen2020}. The gray horizontal lines define the extreme inclinations that delimit the nodal libration region, which depends only on $e_{\text{b}}$ when GR is not considered.

Survival maps with and without GR show the existence of vertical lines of instability, which are the result of mean motion resonances. In addition, it is evident that GR produces dynamical instabilities close to the binary for outer test particles with initial conditions associated with different types of orbits. The instability can reach semimajor axes comparable to and even slightly larger than $a_{\text{lim}}$. For the binary parameters that we have considered, particles may be unstable up to a semimajor axis of about $8\,a_{\rm b}$, a factor of two larger than without the effects of GR. In the next section we consider different regions of instability that are driven by the effects of GR.

\section{Instability driven by GR }
\label{sec:section3}

The effects of GR introduce three strongly unstable regions in the plane ($a_{\text{ini}}$, $i^{\text{e}}_{\text{ini}}$) of significant interest associated with our default scenario:
\begin{enumerate}
    \item Region I (RI) refers to prograde nodal circulating initial orbits. The boundary of RI decreases in inclination $i^{\text{e}}_{\text{ini}}$ with increasing semimajor axis $a_{\text{ini}}$. 
    \item Region II (RII) is associated with nodal librating initial trajectories correlated with both orbital flips and purely retrograde inclinations.
    \item  Region III (RIII) is associated with preferentially retrograde nodal circulating initial orbits for values of $a_{\text{ini}}$ greater than those associated with the minimum extreme inclinations that delimit the nodal libration region.
\end{enumerate}
Figure~\ref{fig:ejemplo_regiones} illustrates the RI, RII, and RIII regions in the plane ($a_{\text{ini}}$, $i^{\text{e}}_{\text{ini}}$) for our default binary parameters.

From Fig.~\ref{fig:mapa_singr}, it is evident that the absence of GR modifies the generation of unstable orbits in the plane ($a_{\text{ini}}$, $i^{\text{e}}_{\text{ini}}$). The unstable regions RI and RIII are not present in the maps that do not include GR. The RII region produced with only classical Newtonian gravity shows a percentage of unstable orbits that is noticeably lower than that obtained with the inclusion of GR. To better understand the features observed in the maps of Figs.~\ref{fig:mapa_gr} and \ref{fig:mapa_singr}, we analyze in detail the dynamics of some case studies of test particles associated with each of those strongly unstable regions produced by GR.

\begin{figure*}[ht!]
\includegraphics[width=\textwidth]{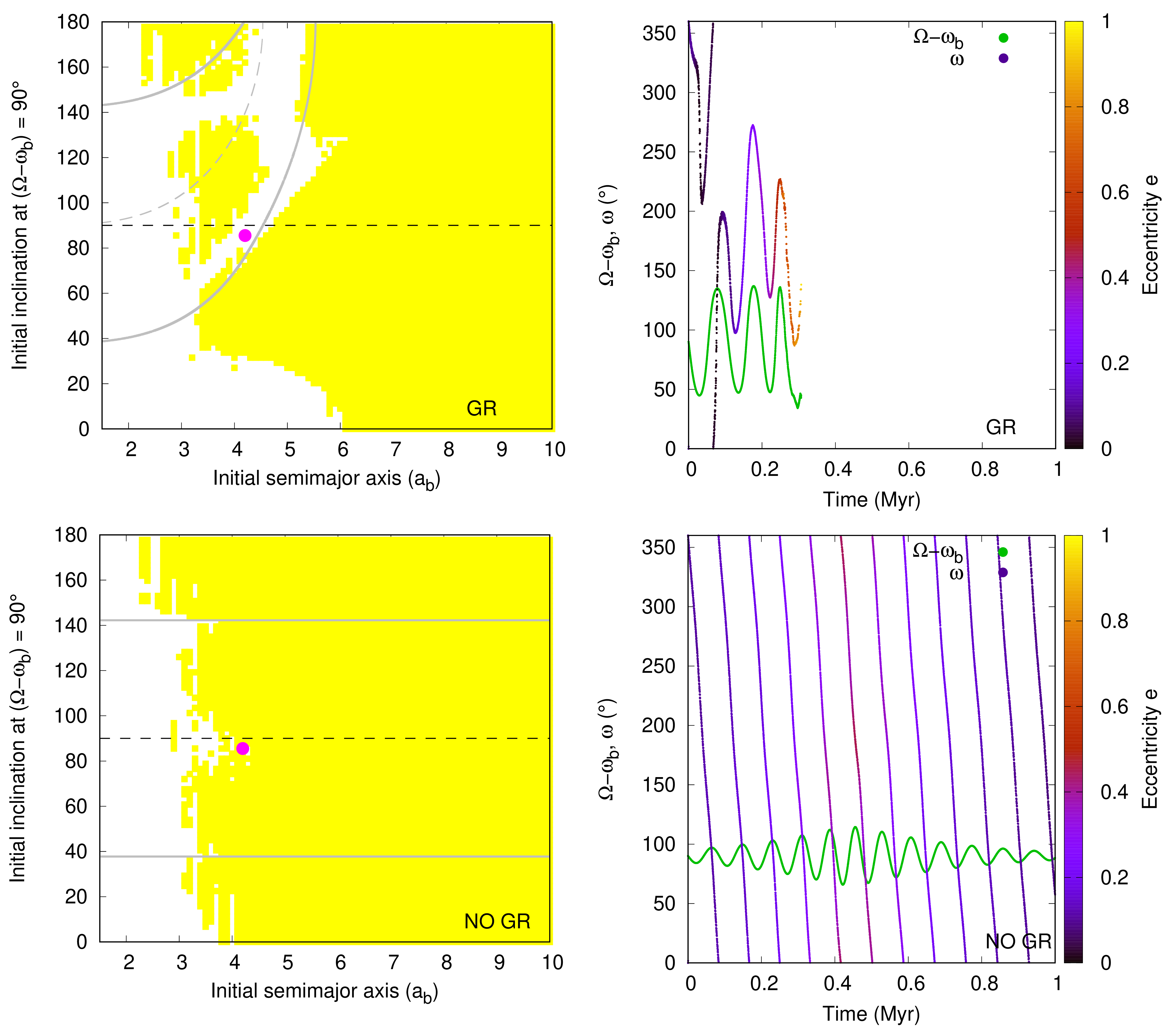}
\caption{As in Fig.~\ref{fig:ejemplo_region2_1}, except that the test particle belonging to the RII region has an initial prograde $i^{\text{e}}$. 
\label{fig:ejemplo_region2_2}}
\end{figure*}

\subsection{Instability in RI}

The top panels of Fig.~\ref{fig:ejemplo_region1} illustrate the particular case of a test particle in RI, which becomes unstable after a time of about $0.3\, $Myr. This particle experiences circulations of $\Omega - \omega_{\text{b}}$ and $\omega$, while $\Omega - \omega_{\text{b}} + \omega$ oscillates and $e$ increases to a value close to 1, promoting instabilities. In the absence of GR, the bottom panels of Fig.~\ref{fig:ejemplo_region1} show that $\Omega - \omega_{\text{b}}$, $\omega$, and $\Omega - \omega_{\text{b}} + \omega$ circulate and $e$ oscillates reaching a maximum value of 0.25, which leads to stable behavior. 

\subsection{Instability in RII}

The top right panels of Figs.~\ref{fig:ejemplo_region2_1} and \ref{fig:ejemplo_region2_2}  describe the temporal evolution of $\Omega - \omega_{\text{b}}$ and $\omega$ of two test particles of RII with different initial conditions associated with $a$ and $i^{\text{e}}$. In both cases, the particle experiences nodal librations and $\omega$ oscillates, which leads to an increase in $e$ to values close to 1. The bottom right panels of Figs~\ref{fig:ejemplo_region2_1} and \ref{fig:ejemplo_region2_2} show that the particles also undergo nodal librations in the absence of GR, although in these cases $\omega$ circulates keeping the value of $e$ bounded over a time of $1 \,$Myr. In fact, $e$ remains below 0.29 and 0.455 for the examples associated with Figs.~\ref{fig:ejemplo_region2_1} and \ref{fig:ejemplo_region2_2}, respectively.

\begin{figure*}[ht!]
\includegraphics[width=\textwidth]{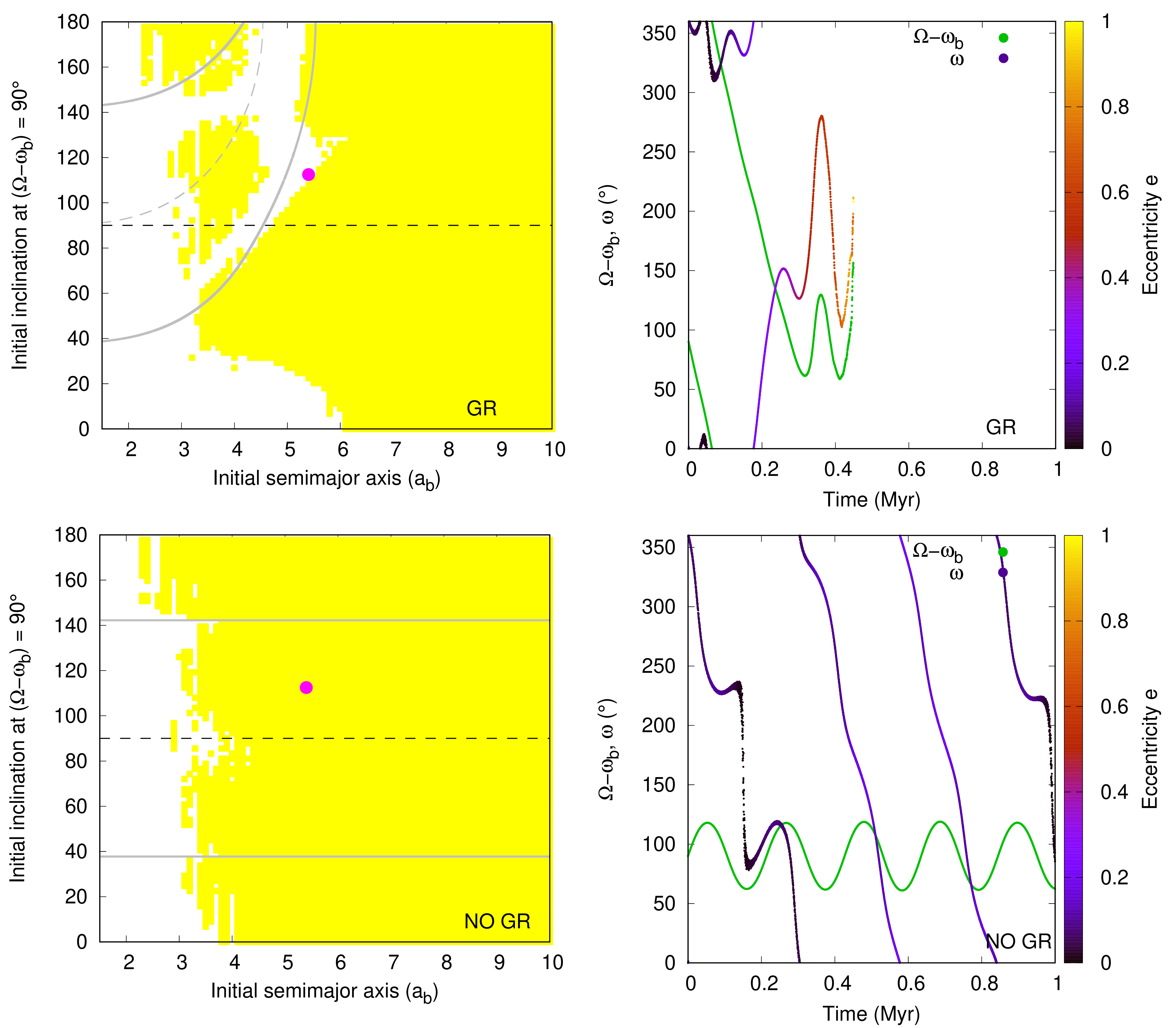}
\caption{As in Fig.~\ref{fig:ejemplo_region2_1}, except that the test particle belongs to the RIII region. 
\label{fig:ejemplo_region3}}
\end{figure*}

\subsection{Instability in RIII}

The top right panel of Fig.~\ref{fig:ejemplo_region3} illustrates the evolution of a test particle in RIII, which is somewhat more complex than in the previous cases. From the beginning, this particle has orbital parameters that lead it to experience nodal circulations. At the same time, $\omega$ oscillates, which leads to growth of $e$. An increase in $e$ leads to an increase in the value of the upper limit of the semimajor axis below which nodal librations of a test particle are possible for suitable values of $i^{\text{e}}$ \citep{Zanardi2023}. Thus, once the value of $e$ reaches 0.25, the test particle starts to experience nodal librations, while $\omega$ oscillates, which leads to a significant increase in $e$ up to a value close to 1. In the absence of GR, the bottom right panel of Fig.~\ref{fig:ejemplo_region3} shows that $\Omega - \omega_{\text{b}}$ oscillates, which is in agreement with the initial orbital parameters of the test particle illustrated in the bottom left panel. In this case, $\omega$ circulates and $e$ oscillates reaching a maximum value of 0.21, which leads to orbital stability. 

\subsection{Effect of $f_{\text{b}}$ and $e_{\text{b}}$ on the instability regions}

From Fig.~\ref{fig:mapa_gr}, we can study the sensitivity of the results of our $n$-body experiments that include GR to the parameters $f_{\text{b}}$ and $e_{\text{b}}$. At fixed $f_{\text{b}}$, the instabilities associated with the RI and RIII regions become more significant with increasing $e_{\text{b}}$. The analysis is somewhat more complex for RII, for which we decide to develop a more detailed description. For $f_{\text{b}}$ = 0.006, almost all pairs ($a_{\text{ini}}$, $i^{\text{e}}_{\text{ini}}$) associated with the nodal libration region produce unstable orbits, although there is a small stable zone associated with retrograde $i^{\text{e}}_{\text{ini}}$ and values of $a_{\text{ini}}$ close to the critical semimajor axis $a_{\text{lim}}$ for $e_{\text{b}}$ = 0.2. For $e_{\text{b}}$ = 0.5, we find very few stable initial conditions within the nodal libration region, which have associated values of $i^{\text{e}}_{\text{ini}}$ close to the maximum extreme inclinations given by the upper gray solid curve. Finally, we distinguish a single point capable of producing orbital stability within the nodal libration region for $e_{\text{b}}$ = 0.8, which has initial values of $a$ and $i^{\text{e}}$ of 1.6 $a_{\text{b}}$ and 114.75$^{\circ}$, respectively.

For $f_{\text{b}} = 0.052$ and $e_{\text{b}} = 0.2$, the unstable RII area extends up to a value of $a_{\text{ini}} = 4.6 \,a_{\text{b}}$ beyond which only stable orbits exist within the nodal libration region. When $e_{\text{b}} = 0.5$, we can find unstable orbits of the RII region up to values of $a_{\text{ini}}$ very close to the critical semimajor axis $a_{\text{lim}}$. Beyond this, this scenario shows three distinctive stable areas within the libration region. On the one hand, a zone with $a_{\text{ini}}$ between 3.4$\,a_{\text{b}}$ and 4.2$\,a_{\text{b}}$ and prograde values of $i^{\text{e}}_{\text{ini}}$ close to the minimum extreme inclinations of the nodal libration region given by the lower gray solid curve. On the other hand, an area with $a_{\text{ini}}$ between 3.1$\,a_{\text{b}}$ and 4.4$\,a_{\text{b}}$ and retrograde values of $i^{\text{e}}_{\text{ini}}$ delimited by the maximum extreme inclinations of the nodal libration region given by the upper gray solid curve. Finally, a wide zone with $a_{\text{ini}}$ ranging from 2.8$\,a_{\text{b}}$ to 4.6$\,a_{\text{b}}$, and prograde and retrograde values of $i^{\text{e}}_{\text{ini}}$. For $e_{\text{b}} = 0.8$, the RII region becomes more strongly unstable and extends to the entire range of $a_{\text{ini}}$ that define the nodal libration region.  Beyond a very small stable area defined by $a_{\text{ini}}$ between 3.4$\,a_{\text{b}}$ and 3.5$\,a_{\text{b}}$ and $i^{\text{e}}$ between 158.75$^{\circ}$ and 168.75$^{\circ}$, there is a very wide stable zone for $a_{\text{ini}}$ ranging between 1.7$\,a_{\text{b}}$ and 4.4$\,a_{\text{b}}$ and prograde and retrograde values of $i^{\text{e}}_{\text{ini}}$. 

For $f_{\text{b}} = 0.24$ and $e_{\text{b}} = 0.2$, a stable zone appears in the libration region from $a_{\text{ini}} = 3.1 \,a_{\text{b}}$, which is surrounded by the unstable region RII up to $a_{\text{ini}} = 5.2 \,a_{\text{b}}$. Beyond this value, there is a very wide zone within the nodal libration region whose pairs ($a_{\text{ini}}$, $i^{\text{e}}_{\text{ini}}$) only produce stable orbits for the test particle. For $e_{\text{b}} = 0.5$, the unstable RII area extends up to $a_{\text{ini}} = 6.9 \,a_{\text{b}}$ along the nodal libration region through two branches whose values of $i^{\text{e}}_{\text{ini}}$ increase with increasing $a_{\text{ini}}$. Beyond $a_{\text{ini}} = 6.9\, a_{\text{b}}$, test particles that experience nodal librations have stable orbits. For $e_{\text{b}}= 0.8$, the RII region becomes more strongly unstable along the two aforementioned branches. Similarly to the previous scenario, stable orbits occur beyond $a_{\text{ini}}= 7 \,a_{\text{b}}$ within the libration region for any value of $a_{\text{ini}}$. 

At fixed $e_{\text{b}}$, the instabilities associated with the RII and RIII regions become more significant with a decrease in $f_{\text{b}}$. An analysis about the sensitivity of the unstable region RI to $f_{\text{b}}$ for a given fixed $e_{\text{b}}$ is somewhat more complex.
In this sense, our results show that the RI region is most strongly unstable for $f_{\text{b}}$ = 0.052, which becomes more evident for $e_{\text{b}}$ values of 0.5 and 0.8. Indeed, for $f_{\text{b}} = 0.052$, the RI instabilities extend up to $i^{\text{e}}_{\text{ini}} = 0^{\circ}$ over the whole range of $a_{\text{ini}}$ values $\lesssim$ 6 $a_{\text{b}}$. For $f_{\text{b}} = 0.006$, RI is strongly unstable up to $i^{\text{e}}_{\text{ini}} = 0^{\circ}$ for $a_{\text{ini}}\lesssim 4$ $a_{\text{b}}$, although in this case there are pairs ($a_{\text{ini}}$, $i^{\text{e}}_{\text{ini}}$) within the region that lead to stable orbits. For $f_{\text{b}} = 0.24$, RI is strongly unstable up to $i^{\text{e}}_{\text{ini}} = 0^{\circ}$ but only for a limited range of $a_{\text{ini}}$ values, which are between 7.4$\,a_{\text{b}}$ and 8$\, a_{\text{b}}$ for $e_{\text{b}}= 0.5$, and between 6.1$\, a_{\text{b}}$ and 7.5$\,a_{\text{b}}$ for $e_{\text{b}}= 0.8$. In fact, in this case, the lower boundary of RI is delimited by a stable region, which reaches values of $i^{\text{e}}_{\text{ini}} = 0^{\circ}$ and extends to an $a_{\text{ini}}$ close to 3.9$\, a_{\text{b}}$ and 4.5$\, a_{\text{b}}$ for $e_{\text{b}}= 0.5$ and 0.8, respectively. It is important to note that there is no RI region for the scenario defined by $f_{\text{b}} = 0.24$ and $e_{\text{b}} = 0.2$.

\subsection{Innermost stable orbits}

We also analyze the sensitivity of the innermost stable orbits to $e_{\text{b}}$ and $f_{\text{b}}$. For $e_{\text{b}}= 0.2$ and 0.5, pairs ($a_{\text{ini}}$, $i^{\text{e}}_{\text{ini}}$) associated with retrograde nodal circulations lead to the innermost stable orbits. In particular, we find that the smaller the value of $f_{\text{b}}$, the smaller the initial semimajor axis of the innermost stable orbit, which is called $a^{\text{iso}}_{\text{ini}}$. Moreover, at a fixed $f_{\text{b}}$, $a^{\text{iso}}_{\text{ini}}$ decreases with decreasing $e_{\text{b}}$. In fact, for $e_{\text{b}} = 0.2$, $a^{\text{iso}}_{\text{ini}}$ has values of 1.5$\,a_{\text{b}}$, 1.7$\,a_{\text{b}}$, and 1.9$\,a_{\text{b}}$ for $f_{\text{b}}= 0.006$, 0.054, and 0.24, respectively, while $a^{\text{iso}}_{\text{ini}}$ increases to 1.7$\, a_{\text{b}}$, 2.3$\,a_{\text{b}}$, and 2.8$\,a_{\text{b}}$ for $f_{\text{b}}= 0.006$, 0.054, and 0.24, respectively, when $e_{\text{b}} = 0.5$. For $e_{\text{b}} = 0.8$, initial conditions associated with nodal librations produce the innermost stable orbits, where the smaller the $f_{\text{b}}$, the smaller the $a^{\text{iso}}_{\text{ini}}$. In this sense, our results show that $a^{\text{iso}}_{\text{ini}}$ has values of 1.6$\,a_{\text{b}}$, 1.7$\,a_{\text{b}}$, and 2.6$\,a_{\text{b}}$ for $f_{\text{b}}= 0.006$, 0.052, and 0.24, respectively. Beyond this, it is worth mentioning that the innermost stable orbit associated with the scenario defined by $f_{\text{b}} = 0.006$ and $e_{\text{b}} = 0.8$ only represents an isolated solution within a very strongly unstable region with nodal librations.

\subsection{Particle removal rate}

\begin{figure*}
\includegraphics[angle=0, width=\textwidth]{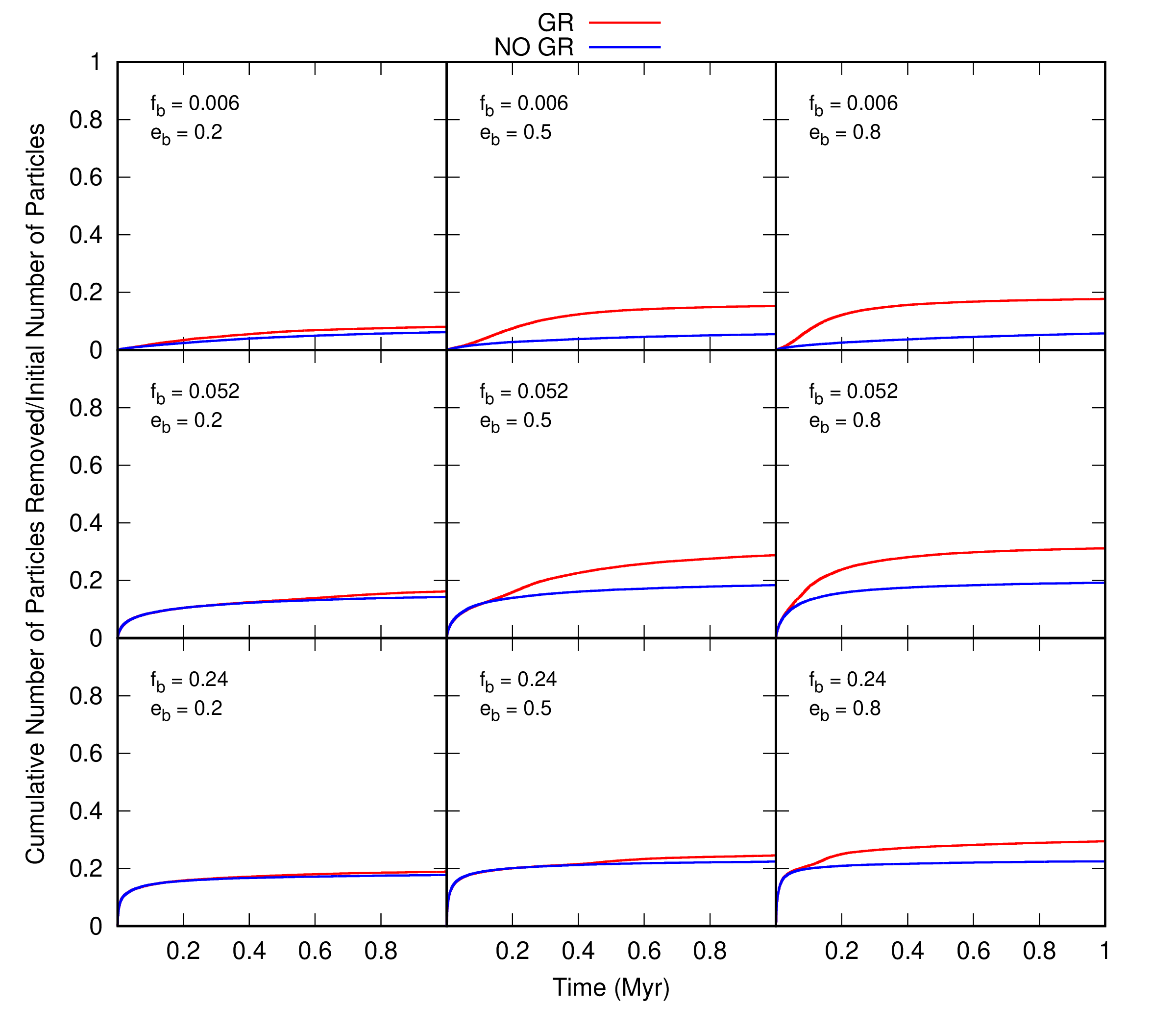}
\caption{Cumulative number of particles removed normalized to the initial
number of particles as a function of time for each of our scenarios of study,
which are defined by $f_{\text{b}}$ and $e_{\text{b}}$. In each panel, the red and blue curves ilustrate the results of our numerical experiments with and without GR, respectively.
\label{fig:Nvst}}
\end{figure*}

Figure~\ref{fig:Nvst} illustrates the cumulative number of particles removed normalized to the total initial number of particles as a function of time for all our scenarios of work. In each panel, the red and blue curves represent the results of simulations with and without GR effects, respectively. From this, several results of interest can be observed.

In general terms, the cumulative fraction of particles removed is higher in simulations that include GR compared to those based on classical Newtonian gravity. For $f_{\text{b}}$ = 0.006 and 0.052, this trend becomes more noticeable for $e_{\text{b}}$ values of 0.5 and 0.8, while for $f_{\text{b}}$ = 0.24, this result is evident mainly for $e_{\text{b}}$ = 0.8. For $e_{\text{b}}$ = 0.2, the cumulative fraction of particles removed in experiments with and without GR is comparable for any $f_{\text{b}}$ value adopted in our study.

On the other hand, for a fixed value of $f_{\text{b}}$, the cumulative fraction of particles removed with GR shows an increasing trend with $e_{\text{b}}$. In the absence of GR, the sensitivity of the cumulative fraction of particles removed to $e_{\text{b}}$ is less noticeable.

Finally, we calculate the particle removal rates over the last 10$^{5}$ years of integration for all the systems under study. At 1 Myr, experiments with GR produce the removal of 1 particle every 452 yr – 1492 yr, while simulations without GR result in the removal of 1 particle every 877 yr - 3225 yr. These results suggest that additional instabilities will occur beyond our integration window at the aforementioned low rates.

\section{Variation of results with simulation parameters}
\label{sec:seccion_testeo_parametros}

We now consider how the results vary with some parameters that we have chosen. We are interested in analyzing variations of the described results for our default scenario ($f_{\text{b}}$ = 0.052 and $e_{\text{b}}$ = 0.5) with parameters such as the semimajor axis $a_{\text{oc}}$ associated with the outer cutoff that defines the removal criterion, and the initial eccentricity $e_{\text{ini}}$ and the initial ascending node longitude $\Omega_{\text{ini}}$ of the outer particles.

\subsection{Dependence on $a_{\text{oc}}$}

Since the analysis developed in the present investigation is based on survival maps, we analyze the sensitivity of the results to the outer cutoff radius, $a_{\rm oc}$, adopted to define the removal criterion for the simulated particles. To do this, we carry out numerical experiments with the aim of studying the dynamical behavior of particles in the default scenario with GR, assuming values of $a_{\text{ini}}$ between 1.5 $a_{\text{b}}$ and 20 $a_{\text{b}}$, and an outer cutoff semimajor axis $a_{\text{oc}}$ = 100 $a_{\text{b}}$. Note that for this new set of simulations, we adopt a wider range of $a_{\text{ini}}$ than in our previous experiments, in order to determine if new instability regions could develop for $a_{\text{ini}} > $10 $a_{\text{b}}$ from a more extended outer cutoff.

The results of our simulations are not sensitive to the outer cutoff semimajor axis for the range of parameters considered. In fact, the survival maps in the plane ($a_{\text{ini}}$, $i^{\text{e}}_{\text{ini}}$) that consider $a_{\text{oc}}$ values of 20 $a_{\text{b}}$ and 100 $a_{\text{b}}$ show very similar dynamical features for $a_{\text{ini}}$ between 1.5 $a_{\text{b}}$ and 10 $a_{\text{b}}$. Moreover, our analysis indicates that a more extended outer cutoff does not lead to new instability regions for $a_{\text{ini}} >$ 10 $a_{\text{b}}$ over the total integration interval of 1 Myr. Indeed, all particles simulated with $a_{\text{ini}}$ values between 10 $a_{\text{b}}$ and 20 $a_{\text{b}}$ survive throughout the entire integration for $a_{\text{oc}} =$ 100 $a_{\text{b}}$.

\begin{figure*}
\includegraphics[angle=270, width=\textwidth]{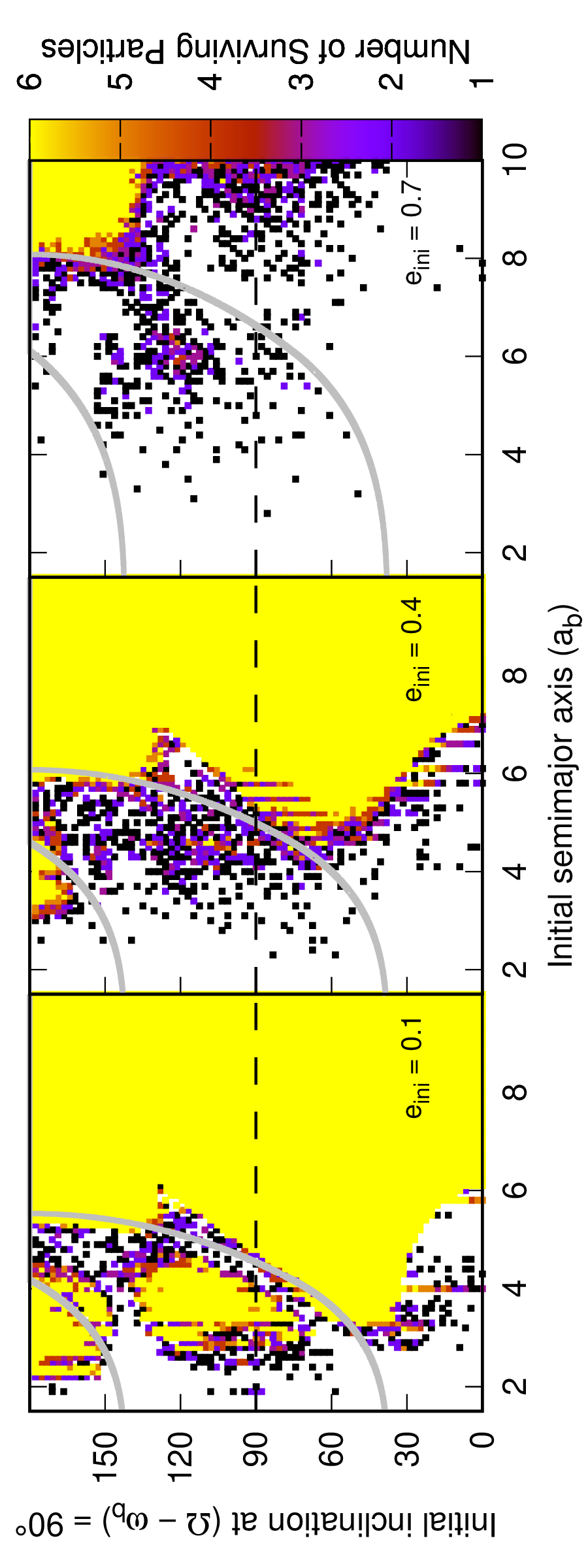}
\caption{Survival maps for our default scenario ($f_{\text{b}}$ = 0.052 and $e_{\text{b}}$ = 0.5) in the
plane ($a_{\text{ini}}$, $i^{\text{e}}_{\text{ini}}$) with GR, assuming $e_{\text{ini}}$ of 0.1 (left panel), 0.4 (middle panel), and 0.7 (right panel). The solid gray curves delimit the nodal libration region. The color code represents the number of
surviving particles after 1 Myr, while the white areas are regions where no particles survive.
\label{fig:grafico_epart}}
\end{figure*}

\subsection{Dependence on $e_{\text{ini}}$}

Understanding how the initial eccentricity $e_{\text{ini}}$ of the outer particles orbiting the SMBH binary affects our results is very interesting for several reasons. On the one hand, an increase in $e_{\text{ini}}$ increases the upper limit of the semimajor axis $a_{\text{lim}}$ below which nodal librations are possible \citep{Zanardi2023}. On the other hand, moderate and high values of $e_{\text{ini}}$ can produce close encounters between the particles and the inner binary over a wider range of $a_{\text{ini}}$ from the beginning of the simulation, promoting strong instabilities.

For this analysis, we develop numerical simulations of our default scenario with GR, assuming an initial eccentricity $e_{\text{ini}}$ for the outer particles of 0.4 and 0.7. Figure~\ref{fig:grafico_epart} shows the resulting survival maps in the plane ($a_{\text{ini}}$, $i^{\text{e}}_{\text{ini}}$) for $e_{\text{ini}}$ values of 0.1 (left panel), 0.4 (middle panel), and 0.7 (right panel). From these panels, it is evident that an increase in $e_{\text{ini}}$ leads to more significant instabilities, decreasing the region of the plane ($a_{\text{ini}}$, $i^{\text{e}}_{\text{ini}}$) associated with the yellow points, which define conditions that lead to all simulated particles to survive throughout the entire integration. In this line of analysis, it is noteworthy how the nodal libration region is devoid of yellow points for the scenarios assuming $e_{\text{ini}}$ = 0.4 and 0.7. In particular, the resulting map for $e_{\text{ini}}$ = 0.7 shows that the only parameters that lead to the survival of all simulated particles throughout the entire integration are restricted to $i^{\text{e}}_{\text{ini}} \gtrsim$ 137$^{\circ}$ and $a_{\text{ini}} >$ $a_{\text{lim}}$.

\subsection{Dependence on $(\Omega -\omega_{\text{b}})_{\text{ini}}$}

\begin{figure}
\includegraphics[angle=270, width=0.5\textwidth]{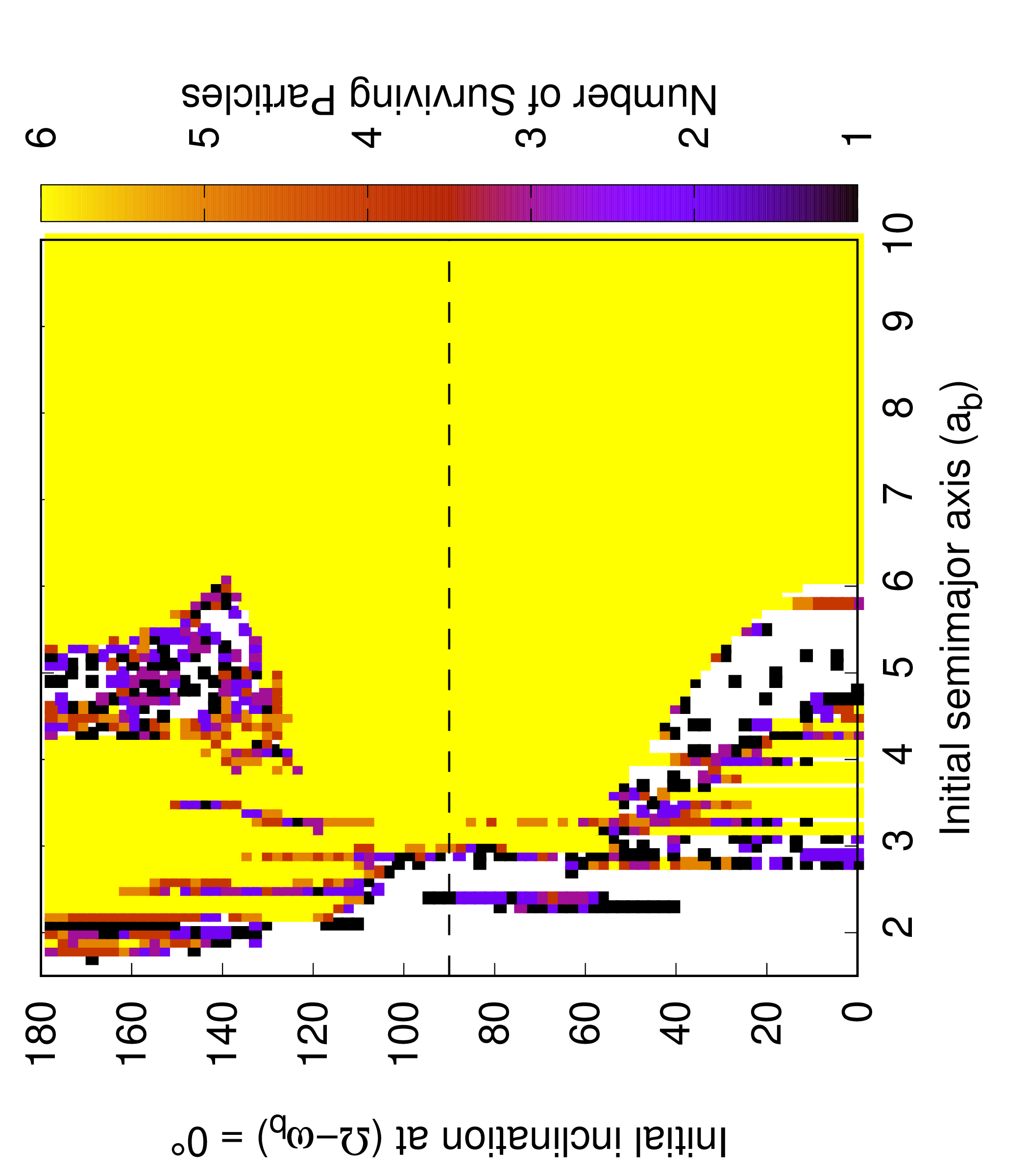}
\caption{Survival map for our default scenario ($f_{\text{b}}$ = 0.052 and $e_{\text{b}}$ = 0.5) in the plane ($a_{\text{ini}}$, $i_{\text{ini}}$) with GR, assuming an initial value of $(\Omega - \omega_{\text{b}}) = 0^{\circ}$. The color code
illustrates the number of surviving particles after 1 Myr. The white areas are regions where no particles survive.
\label{fig:nodo_0}}
\end{figure}

Figure~\ref{fig:nodo_0} illustrates a survival map for our default scenario in the plane ($a_{\text{ini}}$, $i_{\text{ini}}$) with GR effects assuming initial values of $\Omega = \omega = \omega_{\text{b}} = 0^{\circ}$. Since $(\Omega - \omega_{\text{b}})_{\text{ini}} = 0^{\circ}$, it is important to note that the initial conditions of all particles correspond to the nodal circulation regime \citep{Zanardi2018}.

An analysis of this survival map allows us to distinguish two strongly unstable regions. On the one hand, one of them is analogous to the RI region defined in the survival map that assumes $(\Omega - \omega_{\text{b}})_{\text{ini}} = 90^{\circ}$. Indeed, it has associated prograde nodal circulating initial orbits, and the initial inclination that defines its upper limit decreases with an increase in $a_{\text{ini}}$. The dynamical behavior of the particles in this region is very similar to that described for those associated with RI. In fact, while $\Omega$ and $\omega$ circulate throughout the evolution, $\Omega - \omega_{\text{b}} + \omega$ oscillates, increasing the value of $e$, which produces instabilities.

On the other hand, we find a strongly unstable region for $i_{\text{ini}} \gtrsim$ 130$^{\circ}$ and 4 $a_{\text{b}} \lesssim a_{\text{ini}} \lesssim$ 6 $a_{\text{b}}$. In this case, the dynamic evolution of the particles is very similar to that associated with RIII in the survival map with $(\Omega - \omega_{\text{b}})_{\text{ini}}$ = 90$^{\circ}$. Indeed, these particles initially undergo nodal circulations, while $\omega$ oscillates and $e$ increases, which leads to redefining the limits of the nodal libration region. An increase in $e$ decreases the minimum extreme inclination and increases the range of semimajor axes that delimit the nodal libration region \citep{Zanardi2023}. Thus, when $e$ exceeds a certain value, the particle begins to experience nodal librations, while $\omega$ continues to oscillate with increments of $e$, which promotes instabilities.

\section{Discussion and Conclusions} 
\label{sec:section4}

We study the dynamics and stability of low-eccentricity test particles around a supermassive black hole binary system using the {\sc rebound} $n$-body code and the grfull package included in {\sc rebound}x. We analyze the sensitivity of the results to different values of the mass fraction $f_{\text{b}}$ and the eccentricity $e_{\text{b}}$ of the binary. Moreover, we also develop a set of simulations without considering GR effects in order to gain a better understanding of how GR influences the stability of particles in our study systems.

We find that GR promotes strong instabilities around a SMBH binary system, which become evident up to semimajor axes comparable to $a_{\text{lim}}$, both for particles experiencing nodal librations and nodal circulations. In particular, we distinguish three different strongly unstable regions. First, the RI region, which has particles with prograde nodal circulating initial orbits and can experience instabilities at low inclinations, even down to $i^{\text{e}}_{\text{ini}}$ = 0$^{\circ}$. Second, the RII region, whose particles have initial conditions associated with nodal librations. Finally, the RIII region, where particles have preferentially retrograde nodal circulating initial orbits with semimajor axes larger than those defined by the minimum extreme inclinations defining the nodal libration region. We show that significant increases in the orbital eccentricities of particles belonging to the aforementioned regions are correlated with oscillations of $\Omega - \omega_{\text{b}} + \omega$ in RI and of $\omega$ in RII and RIII. Furthermore, we find that the level of instability in such regions is sensitive to $f_{\text{b}}$ and $e_{\text{b}}$. On the one hand, at fixed $f_{\text{b}}$, the larger the $e_{\text{b}}$, the more strongly unstable the RI and RIII regions. On the other hand, at fixed $e_{\text{b}}$, the level of instability in RII and RIII becomes more significant with decreasing $f_{\text{b}}$. Moreover, we observe that the type of trajectory defining the innermost stable orbits depends on $e_{\text{b}}$. Indeed, the innermost stable orbits are the retrograde nodal circulating ones at low and moderate $e_{\text{b}}$, while they are the nodal librating ones at high $e_{\text{b}}$. Finally, we find that instabilities promoted by GR are more significant with an increase in the initial eccentricity of the particles.

It is worth noting that the present investigation only considers first-order post-Newtonian effects to study the dynamics around a SMBH binary system. In fact, we disregard the role of higher-order post-Newtonian effects, both conservative and non-conservative, the latter being responsible for describing the energy losses due to the emission of GWs, which can cause the binary to inspiral and merge \citep{Einstein1918}. According to \cite{Peters1963} and \cite{Peters1964}, the timescale $\tau_{\text{GW}}$ for a GW merger of an isolated binary decreases with an increase in the total mass, the mass fraction, and the eccentricity of the system, and with a decrease in the semimajor axis. From the values adopted for these parameters in the present study, it is possible to verify that the values of $\tau_{\text{GW}}$ are greater than 1 Gyr in all our work scenarios, which are much longer than the total integration time of 1 Myr assumed for our $n$-body experiments.

In this study, we focus on the secular effects responsible for the excitation of the eccentricity and inclination of the particles in the three unstable regions RI, RII, and RIII promoted by GR in the plane ($a_{\text{ini}}$, $i^{\text{e}}_{\text{ini}}$). In addition to these effects, the survival maps illustrated in Figs.~\ref{fig:mapa_gr} and \ref{fig:mapa_singr} show the existence of mean motion resonances that also promote instabilities, mainly for $a_{\text{ini}} \lesssim$ 4 $a_{\text{b}}$. According to works such as \cite{Wisdom1980}, \cite{Mudryk2006}, \cite{Deck2013}, \cite{Ramos2015}, \cite{Hadden2018}, and \cite{Georgakarakos2024}, one key process leading to instability of a three-body system is the overlap of adjacent mean motion resonances. A study concerning the role of GR on circumbinary orbits near mean motion resonances will be a topic of research in future work.

In addition to particle dynamics, our results have implications for circumbinary gas disks.  A circumbinary gas disk undergoes similar dynamics to a test particle. The disk can undergo solid body precession if the communication timescale over the radial extent of the disk is small compared to the precession timescale.  However, viscosity in the disk drives dissipation which leads to disk alignment, either to coplanar to the binary orbit, or polar to the binary orbit \citep{Aly2015,Martin2018}. The end alignment depends upon the type of nodal precession \citep[e.g.][]{Abod2022}. In addition, the effects of GR can therefore change the disk alignment, depending upon the value of $a_{\rm lim}$ and the disk properties \citep{Childs2024}. The inner radius of a circumbinary disk is close to the particle stability limit \citep[e.g.][]{Artymowicz1994,Miranda2015,Franchini2019}.  Our results suggest that the inner radius of a circumbinary disk may be significantly altered by the instabilities driven by GR. The tidal truncation radius of a circumbinary disk can affect the merger timescale of black hole binaries \citep{Heath2020,Lepp2023} and the observable electromagnetic signal from black hole binary mergers \citep{Martin2018b}.

Our results offer important implications for the dynamics and stability around a wide variety of binary systems. 
Polar alignment around highly eccentric main-sequence star binaries and stellar-mass black hole binaries is a likely outcome \citep{Ceppi2024,Johnson2025}. 
While we expect $a_{\rm lim}$ to be tens to hundreds of times the binary separation for a main-sequence star binary \citep{Lepp2022,Zanardi2023}, in a radially extended disk the effects of GR may still be important.  We speculate that unstable particle regions that occur farther out in the disk could lead to disk breaking at these locations.
In view of these results, our study concerning the instability of particles with different inclinations is of significant interest to better understand the global dynamics around binaries on all scales. 

While our research focuses on the dynamics and stability around SMBH binary systems, it would be interesting to analyze whether the strongly unstable regions produced by GR in our study scenarios are present around main-sequence star binaries and stellar-mass black hole binaries. In particular, we consider main-sequence star binaries or even single stars with an inner perturber to be relevant targets for this line of research, as they would lead us to derive important implications for the stability of planets and reservoirs of minor bodies.

Finally, it is important to emphasize that the present research assumes that the outer particles orbiting the SMBH binary system are massless. \citet{Chen2020} analyzed the orbital stability of massive particles around an eccentric main-sequence star binary for different values associated with the physical and orbital parameters of the system. The authors found variations in the stability regions with the mass of the outer particles, which was most evident for low values of the mass fraction and eccentricity of the binary. From this, it would be interesting for future works to study the sensitivity of the results concerning orbital stability to the mass of the outer particles, in order to achieve a better understanding of the general dynamical properties of the different bodies that can be part of a real system.

\begin{acknowledgments}
We thank the anonymous referee for her/his comments and suggestions, which helped us improve the manuscript. We also thank Stephen Lepp for useful conversations. G.C.d.E. and M.Z. acknowledge the partial financial support by Facultad de Ciencias Astronómicas y Geof\'{\i}sicas de la Universidad Nacional de La Plata, and Instituto de Astrof\'{\i}sica de La Plata, for extensive use of their computing facilities. Moreover, G.C.d.E. and M.Z. dedicate this work to Julia and Lourdes, the most wonderful binary system around which they orbit every day.
\end{acknowledgments}

\vspace{5mm}



\bibliography{Paper_deElia}{}
\bibliographystyle{aasjournalv7}

\end{document}